\documentclass[twocolumn,english,american,aps,amsfonts,amsmath,prd,tightenlines,nofootinbib]{revtex4}
\usepackage[T1]{fontenc}
\usepackage[latin1]{inputenc}
\usepackage{amssymb}
\usepackage{amsmath}
\usepackage[dvips,final]{graphicx}

\makeatletter
\usepackage{psfrag}

\usepackage{epsf}

\def\be{\begin{equation}}
\def\ee{\end{equation}}
\def\beq{\begin{equation}}
\def\eeq{\end{equation}}
\def\bea{\begin{eqnarray}}
\def\eea{\end{eqnarray}}
\def\bml{\begin{subequations}}
\def\blea{\bml\begin{eqnarray}}
\def\elea{\end{eqnarray}\end{subequations}}

\usepackage{amsfonts}
\usepackage{amsmath}

\usepackage{babel}
\makeatother
\begin{document}

\title{On adiabatic perturbations in the ekpyrotic scenario}

\selectlanguage{english}

\author{A. Linde,$^{1}$ V. Mukhanov,$^{2,3}$    and A. Vikman$^{3}$}

\date{\today}

\affiliation{$^{1}$Department of Physics, Stanford University, Stanford, CA 94305, USA}

\affiliation{$^{2}$Arnold-Sommerfeld-Center for Theoretical Physics, Department f\"ur Physik,
Ludwig-Maximilians-Universit\"at M\"unchen,
Theresienstr. 37, D-80333, Munich, Germany}

\affiliation{$^{3}$CCPP, New York University, Meyer Hall of Physics,
4 Washington Place, New York, NY 10003,USA}

\selectlanguage{american}
\begin{abstract}
In a recent paper, Khoury and Steinhardt \cite{Khoury:2009my} proposed a
way to generate adiabatic cosmological perturbations with a nearly flat spectrum in a
contracting Universe. To produce these perturbations  they used a regime in which the equation of state exponentially rapidly changed during a short time interval. Leaving aside the singularity problem and the difficult question about the possibility to transmit these perturbations from a contracting Universe to the
expanding phase, we will show that the methods used in \cite{Khoury:2009my} are inapplicable for the description of the cosmological evolution and of the process of generation of perturbations in this scenario.

\end{abstract}
\maketitle

\section{Introduction}\label{intro}

Inflationary theory provides a simple solution to many cosmological problems \cite{Star,Guth,New,Chaot}. It also provides a simple mechanism for the generation of adiabatic perturbations of the metric with flat spectrum, which are responsible for  large-scale structure formation and for the observed CMB anisotropy \cite{Mukh,MukhBook}. Many predictions of inflationary cosmology are already confirmed by observations. There were many attempts to propose an alternative, equally compelling cosmological theory, but so far they have all been unsuccessful.

Nothing illustrates this statement better than the history of the development  of the ekpyrotic/cyclic scenario \cite{KOST}. This scenario was supposed to solve all cosmological problems without using a stage of inflation.
 However, the original version of this scenario did not work. The large mass and entropy of the universe remained unexplained, instead of solving the homogeneity problem this scenario only made it worse, and instead of the big bang expected in  \cite{KOST}, there was a big crunch  \cite{KKL,KKLTS}.

As a result, this scenario was  replaced by the cyclic scenario, which postulated the existence of an infinite number of periods of expansion and contraction of the universe  \cite{cyclic}. However, when this scenario was analyzed,  taking into account the effect of particle production in the early universe,  a very different cosmological regime was found \cite{Felder:2002jk,Linde:2002ws}. In the latest version of this scenario, the homogeneity problem is supposed to be solved by the existence of an infinitely many stages of low-scale inflation, each of which should last at least 56 e-folds \cite{Lehners:2009eg}. 

The most difficult problem facing this scenario is the problem of the cosmological singularity. Originally there was a hope that the cosmological singularity problem will be solved in the context of string theory, but despite the attempts of the best experts in string theory, this problem remains unsolved. A recent attempt to solve this problem was made in the so-called `new ekpyrotic scenario' \cite{Creminelli:2006xe,Buchbinder:2007ad,Creminelli:2007aq,Buchbinder:2007tw}. This scenario involved many exotic ingredients, such as a violation of the null energy condition in a model which combined the ekpyrotic scenario \cite{KOST} with the ghost condensate theory  \cite{Arkani-Hamed:2003uy}. However,  this model  contained ghosts  \cite{Kallosh:2007ad}. The negative sign of the higher derivative term in this scenario makes it hard to construct a ghost-free UV completion of this theory \cite{Kallosh:2007ad}; see also \cite{comment} for a discussion of another attempt to improve the new ekpyrotic scenario.  Even if one eventually succeeds in inventing a UV complete generalization of this theory, such a generalization is expected to obey the null energy condition, in which case the new ekpyrotic scenario will not work anyway.

Leaving all of these unsolved problems aside, the authors of the ekpyrotic scenario continue investigating generation of cosmological perturbations in various versions of this scenario. In inflationary theory, these perturbations are generated after the singularity, but in the ekpyrotic scenario they are supposed to be generated in a collapsing universe before the singularity and preserved on the way through the singularity. Since the singularity problem is not solved, one cannot really say much about it, unless one makes additional assumptions. After a few years of debate, most of the experts agreed, following \cite{Lyth:2001pf}, that adiabatic perturbations with a flat spectrum are not produced in the simplest one-field versions of this theory, contrary to the original claims by the authors of the ekpyrotic/cyclic scenario \cite{KOST,cyclic,Khoury:2001zk}. Therefore more complicated versions of the ekpyrotic scenario   were proposed \cite{Lehners:2007ac}, in which the adiabatic perturbations are produced by a mechanism similar to the inflationary curvaton mechanism  \cite{curv}. The simplest versions of the ekpyrotic scenario of this type were ruled out in \cite{Koyama:2007if}.

That is why it is important to examine the recent paper by Khoury and Steinhardt \cite{Khoury:2009my} who have proposed a new mechanism of generation of adiabatic cosmological perturbations with a nearly flat spectrum at the stage prior to the stage of the ekpyrotic collapse. This is the main goal of our paper. In section \ref{homogen}, we will examine the unperturbed background cosmological solution describing the model of Ref. \cite{Khoury:2009my}. As we will see, this solution requires investigation of the cosmological evolution at super-planckian values of curvature, when the methods used in \cite{Khoury:2009my} are not valid. A similar conclusion will be reached in section \ref{pert} where we will show that the theory of cosmological perturbations used in  \cite{Khoury:2009my} is also not valid.

\section{Background model}\label{homogen}

The model relies on investigation of a contracting universe in a theory with the scalar field potential
\begin{equation}
V=V_{0}\left( 1-e^{-c\phi }\right)\ ,  \label{1}
\end{equation}
where $c\gg 1.$ To simplify a comparison of the results, we will use the
same notations and units $\left( 8\pi G=1\right) $ as in \cite{Khoury:2009my}.  We will start with the homogeneous background solution. Ignoring the
evolution of the universe, the equation of motion for $\phi$ is
\begin{equation}
\ddot{\phi}+cV_{0}e^{-c\phi }= 0 \ .  \label{2}
\end{equation}%
It describes evolution of the field $\phi$ when $t$ slowly increases from large negative $t$. Among other solutions, this equation
has the solution used in  \cite{Khoury:2009my}:
\begin{equation}
\phi = \frac{2}{c}\ln \left(- \sqrt{\frac{V_{0}}{2}}ct\right) \ . \label{3}
\end{equation}%
This solution describes a slowly contracting universe with a nearly
unchanging Hubble parameter $H\approx H_{0} <0.$ One can check that during the pre-ekpyrotic stage, at $t_{\rm begin}<t<t_{\rm end}$, where%
\begin{equation}
t_{\rm begin}\simeq \frac{1}{H_{0}},\text{ \ \ }t_{\rm end}\simeq \frac{1}{H_{0}c^{2}}=%
\frac{t_{\rm begin}}{c^{2}},\text{ \ \ }H_{0}=-\sqrt{\frac{V_{0}}{3}} \ , \label{4}
\end{equation}%
this solution does not receive significant corrections even if one takes into account the contraction of the universe and adds the term $3H\dot \phi$ to equation (\ref{2}). 

During the time interval from $t_{\rm begin}$ to $t_{\rm end}$ (\ref{4}) the scale factor does not decrease significantly and therefore one can set it to be $a\simeq 1.$ Note that the time variable $t$ is negative and its
magnitude changes from a large negative value at the beginning to a small
negative value at the end of this contracting stage. At the same time, the
derivative of the Hubble constant $\dot{H}$ and, correspondingly, the equation
of state $w=p/\rho $ change tremendously during the same time
interval. Indeed, 
\begin{equation}
\dot{H}=-\frac{1}{2}\dot{\phi}^{2}\approx -\frac{2}{c^{2}t^{2}}\ ,  \label{5}
\end{equation}%
hence $\dot{H}_{\rm begin}\simeq H_{0}^{2}/c^{2}$ and $\dot{H}_{\rm end}\simeq
c^{2}H_{0}^{2}.$ For the equation of state we have 
\begin{equation}
1+w=-\frac{2}{3}\frac{\dot{H}}{H^{2}}\approx \frac{1}{H_{0}^{2}c^{2}t^{2}}%
\approx \frac{1}{c^{2}}\left( \frac{t_{\rm begin}}{t}\right) ^{2}\approx c^{2}\left( 
\frac{t_{\rm end}}{t}\right) ^{2} \ . \label{6}
\end{equation}%
At the beginning $\left( 1+w\right) _{\rm begin}\approx c^{-2}\ll 1$ and at the
end $\left( 1+w\right) _{\rm end}\approx c^{2}\gg 1.$

According to \cite{Khoury:2009my}, the stage described above during which the spectrum of
perturbations is generated is followed by a subsequent scaling ekpyrotic stage,
when the Hubble scale evolves as %
\begin{equation}
H\simeq H_{0}\left( \frac{t_{\rm end}}{t}\right) .  \label{7}
\end{equation}%
The authors of \cite{Khoury:2009my} derive the following \textit{necessary condition }which
must be satisfied if we want the perturbations generated at the
pre-ekpyrotic stage to be on the observable scales today (see equation (16)
in \cite{Khoury:2009my}):%
\begin{equation}
\sqrt{V_{0}}\leq 10^{-30}\sqrt{\left\vert H_{\rm end}\right\vert } \ ,  \label{8}
\end{equation}%
where%
\begin{equation}
H_{\rm end}\simeq H_{0}\left( \frac{t_{\rm end}}{t_{\rm ekp}}\right)\ ,  \label{9}
\end{equation}%
is the Hubble constant not at $t = t_{\rm end}$, but at the end of the ekpyrotic stage, at  $t = t_{\rm ekp}.$ Taking
into account that $H_{0}\simeq -\sqrt{V_{0}}$ we obtain from (\ref{8}) that
this necessary condition can be rewritten as%
\begin{equation}
\left\vert t_{\rm ekp}\right\vert \leq 10^{-60}H_{0}^{-1}\left\vert
t_{\rm end}\right\vert \simeq 10^{-60}\frac{1}{c^{2}H_{0}^{2}} \ . \label{10}
\end{equation}%
Here time is expressed in units of Planck time $t_{P} \sim M_{P}^{-1} \sim 10^{{-43}}$ s. On the other hand, according to \cite{Khoury:2009my} (see also below), the amplitude of the
generated perturbations is $O(cH_{0})$. Since this amplitude should be $%
O(10^{-5})$, the ekpyrotic stage should end at 
\begin{equation}
\left\vert t_{\rm ekp}\right\vert <10^{-50}\, t_{P} \ . \label{shorttime}
\end{equation}%
In other words, we are dealing here with frequencies which are 50 orders of magnitude greater than the Planck mass! This completely invalidates the use of the classical equations of motion in the investigation of the background solution in \cite{Khoury:2009my}. Moreover, any mechanism addressing the singularity problem (if this mechanism exists at all) should operate on this time scale.  
 
This fact was not noticed in \cite{Khoury:2009my} because the authors concentrated on the energy density at the end of the ekpyrotic stage,
\begin{equation}
\rho \simeq H^{2}\simeq \frac{4}{c^{4}t_{\rm ekp}^{2}}\simeq \frac{%
10^{100}}{c^{4}}\ ,  \label{11}
\end{equation}%
in Planck units.   They studied two different regimes, $c = 10^{28}$ and $c = 10^{40}$. In both cases $\rho \ll 1$, so one could think that we are safely in the sub-Planckian regime. However, at the ekpyrotic stage with $c^{2 } \gg 1$ one has $p \sim c^{2}\rho \gg \rho$, so from the Einstein equation $\dot H = -(\rho + p)/2$ it follows that
\begin{equation}
p \approx -2\dot H \sim c^{2} H^{2}\simeq \frac{4}{c^{2}t_{\rm ekp}^{2}}\simeq \frac{%
10^{100}}{c^{2}} \ , \label{11a}
\end{equation}%
i.e. pressure $p$,  as well as $\dot H$ and the curvature $R$, are more than 20 orders of magnitude greater that the Planck density for both of the values of the parameters discussed in  \cite{Khoury:2009my}.

The situation with higher-order curvature invariants is even worse. For instance, 
\begin{equation*}
\sqrt{\left| R_{;\alpha }R^{;\alpha }\right|}\sim \ddot{H}\sim \frac{1}{c^{2}t_{\rm ekp}^{3}}%
\sim \frac{10^{150}}{c^{2}} \ .
\end{equation*}%
This means that they exceed the Planckian values by 94 orders of magnitude for $c = 10^{28}$, and by 70 orders of magnitude for $c = 10^{40}$. If one considers even higher invariants, the situation becomes more and more troublesome. This is a direct consequence of the abnormal smallness of the time at the end of the stage of the ekpyrosis (\ref{shorttime}), which happens in this scenario for all values of its parameters.

One can easily check that the time $t_{\rm end}$, when the pre-ekpyrotic stage ends and the stage of the ekpyrosis begins in this scenario, is also many orders of magnitude smaller than the Planck time, for the values of the parameters used in   \cite{Khoury:2009my}. Moreover, one can show that this conclusion is valid for all possible values of parameters compatible with the requirement $cH_{0} \sim 10^{{-5}}$ unless there is a long stage of inflation after the ekpyrotic big crunch.

Note that we are not talking here about hypothetical processes which are supposed to lead to a bounce and prevent the cosmological singularity in this scenario. The pre-ekpyrotic stage, as well as the ekpyrotic stage, are described in \cite{Khoury:2009my} by standard methods of classical field theory and general theory of relativity, which are not valid at $|t|< t_{P}$ and at super-planckian values of curvature. Our results  show therefore that the suggested scenario is completely unreliable even as a background model. In the next section we will show that even if one ignores all of the problems discussed above, the calculation of the spectrum of metric perturbations in \cite{Khoury:2009my} is also unreliable.

\section{Perturbations}\label{pert}

The perturbed homogeneous flat universe in the conformal-Newtonian coordinate
system is described by the metric%
\begin{equation}
ds^{2}=a^{2}\left( \eta \right) \left[ \left( 1+2\Phi \right) d\eta
^{2}-\left( 1-2\Phi \right) \delta _{ik}dx^{i}dx^{k}\right] ,  \label{13}
\end{equation}%
where $\Phi $ is the gravitational potential induced by the perturbations of the
scalar field $\delta \phi .$  Linear perturbation theory is reliable only
if $\Phi $, as well as the energy density perturbations $\delta T_{0}^{0}/T_{0}^{0}$,
are simultaneously much smaller than unity. Otherwise 
the higher-order terms in perturbation theory dominate over linear terms, and 
the perturbative expansion fails.

Before going any further, let us illustrate the statements made above using inflationary perturbations as  an example. During inflation, the large-scale energy density perturbations and perturbations of metric are small $\delta T_{0}^{0}/T_{0}^{0} \sim -2\Phi < 10^{-5}$. 
  
 At the transition to the post-inflationary stage,  the value of $\Phi$ grows to about $10^{-5}$ and stabilizes at this level. Meanwhile the energy density perturbations $\delta T_{0}^{0}/T_{0}^{0}$ continue to grow. Once these perturbations become large,  $\delta T_{0}^{0}/T_{0}^{0} \sim O(1)$, galaxies begin to form and separate form the cosmological expansion. The subsequent evolution of the large-scale structure of the universe, {\it as well as the related evolution of $\Phi$}, cannot be described by perturbation theory. However, this happens long after the end of inflation, which is why the standard theory of generation of inflationary perturbations is reliable.

Ref. \cite{Khoury:2009my} did not contain any investigation of the perturbations $\delta T_{0}^{0}/T_{0}^{0}$, which is necessary to examine validity of their results. As we will see, in their model, the perturbations $\delta T_{0}^{0}/T_{0}^{0}$ are large already at the pre-ekpyrotic stage, and therefore one cannot apply perturbation theory for the investigation of the generation of perturbations in the metric and in the energy density in this scenario.

To quantize scalar perturbations one introduces the canonical variable \cite{MukhBook}%
\begin{equation}
v=a\left( \delta \phi +\sqrt{3\left( 1+w\right) }\Phi \right) ,  \label{14}
\end{equation}%
which for every Fourier mode $k$ satisfies the equation%
\begin{equation}
v_{k}^{\prime \prime }+\left( k^{2}-\frac{z^{\prime \prime }}{z}\right)
v_{k}=0 \ ,  \label{15}
\end{equation}%
where prime means the derivative with respect to conformal time $\eta $, and $z=a%
\sqrt{3\left( 1+w\right) }.$ In the case under consideration 
\begin{equation}
\frac{z^{\prime \prime }}{z}\simeq \frac{2}{t^{2}} \ , \label{16}
\end{equation}%
during the time interval $t_{\rm begin}<t<t_{\rm end}.$  During the pre-ekpyrotic stage $%
a\approx 1$, but the second derivative of $a$ can be large and hence
before identifying conformal and physical time we have to perform all
differentiation in our equations, and only after that it is safe to set $%
t\approx \eta $. \ For the short-wavelength perturbations with $k\left\vert
t\right\vert \gg 1$%
\begin{equation}
v_{k}\simeq \frac{1}{\sqrt{k}}e^{ik\eta }\simeq \frac{1}{\sqrt{k}}e^{ikt} \ ,
\label{18}
\end{equation}%
where the amplitude is fixed by the requirement that the perturbations are
the minimal vacuum fluctuations. In the long-wave limit for $k\left\vert
t\right\vert \ll 1$ the solution of (\ref{15}) is%
\begin{equation}
v_{k}\simeq C_{1}z+C_{2}z\int \frac{d\eta }{z^{2}}\simeq \frac{1}{\sqrt{k}}%
\frac{t_{k}}{t} \ ,  \label{19}
\end{equation}%
where we have fixed the constant of integration requiring the
continuity of the solution at the moment of ``effective horizon'' crossing, $%
t_{k}\simeq 1/k$, and have neglected the decaying mode. It follows from here
that for the long-wavelength perturbations the ``conserved'' $\zeta $ is equal%
\begin{equation}
\zeta _{k}\equiv \frac{v_{k}}{z}\simeq \frac{c}{\sqrt{k}}\frac{t_{k}}{t_{b}}%
\simeq \frac{cH_{0}}{k^{3/2}} \ .  \label{20}
\end{equation}%
Therefore at the end of the pre-ekpyrotic stage the spectrum 
\begin{equation}
\zeta _{k}k^{3/2}\simeq cH_{0}   \label{21}
\end{equation}%
is scale-invariant in the range of wavelengths $\left\vert
t_{\rm end}\right\vert <\lambda \,<\left\vert t_{\rm begin}\right\vert$. For $%
cH_{0}\simeq 10^{-5}$ the amplitude of perturbations is of the right
order of magnitude. This result was the reason for the claim in \cite{Khoury:2009my} that
the pre-ekpyrotic stage can produce the perturbations similar to those produced
by inflation. In this case the flat spectrum appears because of the
very fast change of the equation of state in a contracting universe. 

Let us
find under which conditions the result obtained is valid at the end of
the pre-ekpyrotic stage. With this purpose we will calculate the
gravitational potential and energy density perturbations. It is convenient
to introduce the variable $u$ related to the gravitational potential as%
\begin{equation}
\Phi =H\left( 1+w\right) ^{1/2}u \ , \label{22}
\end{equation}%
and satisfying the equation 
\begin{equation}
u_{k}^{\prime \prime }+\left( k^{2}-\frac{\left( 1/z\right) ^{\prime \prime }%
}{1/z}\right) u_{k}=0 \ . \label{23}
\end{equation}%
During the pre-ekpyrotic stage 
\begin{equation}
\frac{\left( 1/z\right) ^{\prime \prime }}{1/z}\simeq -\frac{H_{0}}{t} \ .
\label{24}
\end{equation}%
Therefore, for perturbations with $k^{2}\left\vert t\right\vert \gg
\left\vert H_{0}\right\vert $, we have 
\begin{equation}
u_{k}\propto e^{\pm ikt},  \label{25}
\end{equation}%
while for $k^{2}\left\vert t\right\vert \ll \left\vert H_{0}\right\vert$  the solution is 
\begin{equation}
u_{k}\simeq C_{1}+C_{2}t  \ . \label{26}
\end{equation}%

Using the $0-i$ Einstein equation, 
\begin{equation}
\dot{\Phi}+H\Phi =\frac{1}{2}\dot{\phi}\delta \phi =\frac{\sqrt{3}}{2}%
H\left( 1+w\right) ^{1/2}\delta \phi \ , \label{27}
\end{equation}%
to express $\delta \phi $ in terms of $\Phi $ in (\ref{14}), we obtain \cite{MukhBook}
\begin{equation}
\zeta =\frac{v}{z}=\frac{2}{3}\frac{\dot{\Phi}+H\Phi }{H\left( 1+w\right) }%
+\Phi ,  \label{28}
\end{equation}%
where the dot denotes the derivative with respect to the physical time 
$t=\int ad\eta$. By substituting (\ref{22}) and using the background
equations of motion, according to which%
\begin{equation}
\frac{\dot{H}}{H^{2}}=-\frac{3}{2}\left( 1+w\right) ,  \label{17}
\end{equation}%
we find%
\begin{equation}
\zeta =\frac{2}{3}\left( \frac{1}{2}\frac{\dot{w}}{H\left( 1+w\right) ^{2}}+%
\frac{\dot{u}}{H\left( 1+w\right) u}+\frac{1}{1+w}\right) \Phi. \label{29}
\end{equation}%
It follows from here that%
\begin{equation}
\zeta \simeq -\frac{2}{3}\frac{t}{t_{\rm end}}\Phi  \ , \label{30}
\end{equation}%
for the perturbations with $k\left\vert
t\right\vert \ll 1$ during the pre-ekpyrotic stage. For deriving this result we used (\ref%
{25}) and (\ref{26}). Thus at the end of the pre-ekpyrotic stage $\zeta
\simeq \Phi $ and hence the spectrum of the gravitational potential is the
same as that for $\zeta .$ 

The next thing to do is to determine the amplitude
of the energy density perturbations and verify that they also remain small.
Using the $0-0$ Einstein equation, we find 
\begin{eqnarray}
\left(\frac{\delta T_{0}^{0}}{T_{0}^{0}}\right)_k =-\frac{2k^2}{3H^{2}a^{2}} \Phi_k -%
\frac{2}{H}\left( \dot{\Phi}_k+H\Phi_k \right) = \notag \\
\Bigl(-\frac{2k^{2}}{3H^{2}a^{2}}+ 3\left(1+w\right)-2-\frac{\dot{w}}{%
H\left( 1+w\right) }-\frac{2\dot{u}}{Hu}\Bigr) \Phi_k \ . ~~~\label{31}
\end{eqnarray}%
During the pre-ekpyrotic stage,  this relation for perturbations with $k>H_{0}$ reduces to 
\begin{equation}
\left(\frac{\delta T_{0}^{0}}{T_{0}^{0}}\right)_k \simeq \left( -\frac{2k^{2}}{3H^{2}}+\frac{%
2}{Ht}\right) \Phi_k \ .  \label{32}
\end{equation}%
This equation implies that at the end of the pre-ekpyrotic stage, at $t\simeq t_{\rm end}$%
\begin{equation}
\frac{\delta T_{0}^{0}}{T_{0}^{0}}\simeq 2c^{2}\Phi   \label{33}
\end{equation}%
for $H_{0}< k\ll cH_{0}$ and%
\begin{equation}
\left(\frac{\delta T_{0}^{0}}{T_{0}^{0}}\right)_k\simeq c^{2}\Phi_k \left( \frac{k}{cH_{0}}\right)
^{2} \gg 2c^{2}\Phi_k \ ,  \label{34}
\end{equation}%
for $k\gg cH_{0}.$  

Thus we see that for the observed value $\Phi \sim 3 \times 10^{-5}$ the perturbations of the energy
density exceed unity unless $c\lesssim 10^{2}.$ Note that the large $c$  approximation used in  \cite{Khoury:2009my} requires $c\gg 1$.  Eq. (\ref{34}) implies that even in the marginal case  $1 \ll c\lesssim 10^{2}$ the energy density perturbations exceed unity at all scales smaller than $O(10^{-2}) H_{0}^{-1}$ for $c\lesssim 10^{2}.$  For $c \gtrsim 10^{2}$ we expect that the linear perturbation theory breaks down at all interesting scales.
Let us show that this is what really happens in the case under consideration.

From (\ref{27})
we find that during the pre-ekpyrotic stage
\begin{equation}
\delta \phi \simeq c\Phi  \ ,  \label{35}
\end{equation}%
for perturbations with $k\left\vert t\right\vert \ll 1.$ Expansion
of the energy momentum tensor around the background contains terms of all
orders in $\delta \phi .$ In particular, in the expansion of $\delta T_{0}^{0}
$ we have 
\begin{equation}
\delta T_{0}^{0}=V_{,\phi }\delta \phi +\frac{1}{2}V_{,\phi \phi }\delta
\phi ^{2}+...  \label{36}
\end{equation}%
It is clear that linear perturbation theory is applicable only if the
quadratic term here is small compared to the linear term. Let us find when this may happen. Taking into account that in the case under consideration $V_{,\phi \phi }=cV_{,\phi }$ we find that 
\begin{equation}
\frac{V_{,\phi \phi }\delta \phi ^{2}}{V_{,\phi }\delta \phi }=c\delta \phi
\simeq c^{2}\Phi  \ . \label{37}
\end{equation}%
This ratio does not exceed unity only if $1 \ll c \lesssim 10^{2}$,  in agreement with the conclusion which we just obtained by a different method. Thus  linear
perturbation theory is not applicable for $c \gtrsim 10^{2},$ and it  fails completely for $c = 10^{28}$ and for $c = 10^{40}$ studied in \cite{Khoury:2009my}.

Note that this problem is much more severe than the potentially curable problem of anomalously large nongaussianity discussed in \cite{Khoury:2009my}. The nongaussianity problem appears only for extremely small wavelengths $k^{{-1}} > c^{{-1}}H_{0}^{{-1}}$, where $c$ can be as large as $10^{28}$ or $10^{40}$. Meanwhile for $c \gtrsim  10^{2}$ the problems discussed in our paper occur at all wavelengths. 

Moreover, as we already mentioned, the spectrum is not satisfactory even in the marginal case $1 \ll c \lesssim 10^{2}$.  Eq. (\ref{34}) implies, in particular, that in the limiting case of $c = O(1)$, when the large $c$ approximation used in \cite{Khoury:2009my} breaks down, the spectrum of perturbations of density is blue with $n_{s} \sim 5$, and its amplitude becomes greater than unity on scales two orders of magnitude smaller than $H_{0}^{-1}$. In addition, according to (\ref{11}), in this case one would be forced to consider the ekpyrotic stage ending at the density exceeding the Planck density by 100 orders of magnitude.

\section{Conclusions}

 As we have seen, the new version of the ekpyrotic scenario requires calculations to be performed at  curvatures which are exponentially greater than the Planck curvature. Moreover, the theory of generation of metric perturbations used in this scenario is based on calculations which are exponentially far away from the domain of applicability of this theory. These problems appear independently of the major problem of this scenario, which is the problem of the cosmological singularity.

Each time when a new version of the ekpyrotic/cyclic  theory is proposed, the authors suggest that its predictions be compared with predictions of inflationary cosmology to check which of the theories is better with respect to observations. In accordance with this tradition, the authors of Ref. \cite{Khoury:2009my} say that their new theory ``predicts non-gaussianity and a spectrum of gravitational waves that is observationally distinguishable from inflation.''  The main problem with this suggestion is that one can test any theory only if it is internally consistent and only if it really makes definite predictions.

\

\section*{Acknowledgments}

We are thankful to  Ignacy Sawicki for useful comments. The work of A.L. was supported in part by NSF grant PHY-0244728, 
by the Alexander-von-Humboldt Foundation, and  by the FQXi grant RFP2-08-19. V.M. is supported by  TRR 33 ``The Dark Universe'' and the Cluster of Excellence EXC 153 ``Origin and 
Structure of the Universe''. A.V. is supported by the James Arthur Fellowship.


\begin{thebibliography}{10}


\bibitem{Khoury:2009my}
  J.~Khoury and P.~J.~Steinhardt,
``Adiabatic Ekpyrosis: Scale-Invariant Curvature Perturbations from a Single
Scalar Field in a Contracting Universe,''
  arXiv:0910.2230 [hep-th].
  


\bibitem{Star} A.~A.~Starobinsky, ``A
New Type Of Isotropic Cosmological Models Without Singularity,''
Phys.\ Lett.\ B {\bf 91}, 99 (1980).

\bibitem{Guth} A.~H.~Guth,
``The Inflationary Universe: A Possible Solution To The Horizon
And Flatness Problems,'' Phys.\ Rev.\ D {\bf 23}, 347 (1981).

\bibitem{New}
A.~D.~Linde, ``A New Inflationary Universe Scenario: A Possible
Solution Of The Horizon, Flatness, Homogeneity, Isotropy And
Primordial Monopole Problems,'' Phys.\ Lett.\ B {\bf 108}, 389
(1982).

\bibitem{Chaot} A.~D.~Linde,
``Chaotic Inflation,'' Phys.\ Lett.\ B {\bf 129}, 177 (1983).

\bibitem{Mukh} V.~F.~Mukhanov and G.~V.~Chibisov,
``Quantum Fluctuation And `Nonsingular' Universe,'' JETP Lett.\
{\bf 33}, 532 (1981) [Pisma Zh.\ Eksp.\ Teor.\ Fiz.\  {\bf 33},
549 (1981)].


\bibitem{MukhBook} V. F. Mukhanov, {\it Physical Foundations of Cosmology}, Cambridge University Press, 2005.




  
  
  
  \bibitem{KOST}
J.~Khoury, B.~A.~Ovrut, P.~J.~Steinhardt and N.~Turok, ``The
ekpyrotic universe: Colliding branes and the origin of the hot big
bang,'' Phys.\ Rev.\ D {\bf 64}, 123522 (2001)
[arXiv:hep-th/0103239].



\bibitem{KKL}
R.~Kallosh, L.~Kofman and A.~D.~Linde, ``Pyrotechnic universe,''
Phys.\ Rev.\ D {\bf 64}, 123523 (2001) [arXiv:hep-th/0104073].



\bibitem{KKLTS}
R.~Kallosh, L.~Kofman, A.~D.~Linde and A.~A.~Tseytlin, ``BPS
branes in cosmology,'' Phys.\ Rev.\ D {\bf 64}, 123524 (2001)
[arXiv:hep-th/0106241].



\bibitem{cyclic}
P.~J.~Steinhardt and N.~Turok, ``Cosmic evolution in a cyclic
universe,'' Phys.\ Rev.\ D {\bf 65}, 126003 (2002)
[arXiv:hep-th/0111098].

\bibitem{Felder:2002jk}
  G.~N.~Felder, A.~V.~Frolov, L.~Kofman and A.~V.~Linde,
``Cosmology with negative potentials,''
  Phys.\ Rev.\  D {\bf 66}, 023507 (2002)
  [arXiv:hep-th/0202017].

\bibitem{Linde:2002ws}
  A.~Linde,
``Inflationary theory versus ekpyrotic/cyclic scenario,'' in {\it The future of theoretical physics and cosmology}, (Cambridge Univ. Press, Cambridge 2002), p. 801
  [arXiv:hep-th/0205259].
  
\bibitem{Lehners:2009eg}
  J.~L.~Lehners, P.~J.~Steinhardt and N.~Turok,
``The Return of the Phoenix Universe,''
  arXiv:0910.0834 [hep-th].




\bibitem{Creminelli:2006xe}
  P.~Creminelli, M.~A.~Luty, A.~Nicolis and L.~Senatore,
   ``Starting the universe: Stable violation of the null energy condition and
non-standard cosmologies,''
  JHEP {\bf 0612}, 080 (2006)
  [arXiv:hep-th/0606090].


\bibitem{Buchbinder:2007ad}
  E.~I.~Buchbinder, J.~Khoury and B.~A.~Ovrut,
``New Ekpyrotic Cosmology,''
  Phys.\ Rev.\  D {\bf 76}, 123503 (2007)
  [arXiv:hep-th/0702154].

\bibitem{Creminelli:2007aq}
  P.~Creminelli and L.~Senatore,
``A smooth bouncing cosmology with scale invariant spectrum,''
  JCAP {\bf 0711}, 010 (2007)
  [arXiv:hep-th/0702165].

\bibitem{Buchbinder:2007tw}
  E.~I.~Buchbinder, J.~Khoury and B.~A.~Ovrut,
``On the Initial Conditions in New Ekpyrotic Cosmology,''
  JHEP {\bf 0711}, 076 (2007)
  [arXiv:0706.3903 [hep-th]].


\bibitem{Arkani-Hamed:2003uy}
  N.~Arkani-Hamed, H.~C.~Cheng, M.~A.~Luty and S.~Mukohyama,
``Ghost condensation and a consistent infrared modification of gravity,''
  JHEP {\bf 0405}, 074 (2004)
  [arXiv:hep-th/0312099].
  
\bibitem{Kallosh:2007ad}
  R.~Kallosh, J.~U.~Kang, A.~D.~Linde and V.~Mukhanov,
 ``The New Ekpyrotic Ghost,''
  JCAP {\bf 0804}, 018 (2008)
  [arXiv:0712.2040 [hep-th]].
  
  \bibitem{comment}
It was suggested in \cite{Creminelli:2008wc} that one might avoid the problems of the new ekpyrotic scenario discussed in \cite{Kallosh:2007ad} by using an approach to the effective field theory with higher derivatives proposed in \cite{Weinberg:2008hq}. Ref. \cite{Weinberg:2008hq} suggested to  use iterations, initially solving equations of the theory omitting the higher derivative terms, and then substituting the solutions to the higher derivative terms. This proposal works for a large class of theories where omission of the higher-derivative terms does not lead to pathologies. However, it does not help in the case of the new ekpyrotic scenario because this scenario, with the higher derivative being omitted, exhibits a catastrophic gradient instability \cite{Kallosh:2007ad}. 

\bibitem{Creminelli:2008wc}
  P.~Creminelli, G.~D'Amico, J.~Norena and F.~Vernizzi,
``The Effective Theory of Quintessence: the w<-1 Side Unveiled,''
  JCAP {\bf 0902}, 018 (2009)
  [arXiv:0811.0827 [astro-ph]].

  
  
\bibitem{Weinberg:2008hq}
  S.~Weinberg,
``Effective Field Theory for Inflation,''
  Phys.\ Rev.\  D {\bf 77}, 123541 (2008)
  [arXiv:0804.4291 [hep-th]].

  

\bibitem{Lyth:2001pf}
  D.~H.~Lyth,
``The primordial curvature perturbation in the ekpyrotic universe,''
  Phys.\ Lett.\  B {\bf 524}, 1 (2002)
  [arXiv:hep-ph/0106153];
  D.~H.~Lyth,
``The failure of cosmological perturbation theory in the new ekpyrotic
scenario,''
  Phys.\ Lett.\  B {\bf 526}, 173 (2002)
  [arXiv:hep-ph/0110007].


\bibitem{Khoury:2001zk}
  J.~Khoury, B.~A.~Ovrut, P.~J.~Steinhardt and N.~Turok,
``Density perturbations in the ekpyrotic scenario,''
  Phys.\ Rev.\  D {\bf 66}, 046005 (2002)
  [arXiv:hep-th/0109050].



\bibitem{Lehners:2007ac}
  J.~L.~Lehners, P.~McFadden, N.~Turok and P.~J.~Steinhardt,
``Generating ekpyrotic curvature perturbations before the big bang,''
  Phys.\ Rev.\  D {\bf 76}, 103501 (2007)
  [arXiv:hep-th/0702153];
A.~J.~Tolley and D.~H.~Wesley,
``Scale-invariance in expanding and contracting universes from two-field
models,''
  JCAP {\bf 0705}, 006 (2007)
  [arXiv:hep-th/0703101].
  

  \bibitem{curv}  A.~D.~Linde and V.~Mukhanov,  ``Nongaussian isocurvature
perturbations from inflation,''  Phys.\ Rev.\ D \textbf{56}, 535 (1997)
[arXiv:astro-ph/9610219]; K.~Enqvist and M.~S.~Sloth,
``Adiabatic CMB perturbations in pre big bang string cosmology,''
  Nucl.\ Phys.\  B {\bf 626}, 395 (2002)
  [arXiv:hep-ph/0109214]; D.~H.~Lyth and D.~Wands, ``Generating the curvature
perturbation without an inflaton,''  Phys.\ Lett.\ B \textbf{524}, 5 (2002)
[arXiv:hep-ph/0110002];   T.~Moroi and T.~Takahashi, ``Effects of cosmological
moduli fields on cosmic microwave background,''  Phys.\ Lett.\ B \textbf{522}, 215 (2001) [arXiv:hep-ph/0110096].




\bibitem{Koyama:2007if}
  K.~Koyama, S.~Mizuno, F.~Vernizzi and D.~Wands,
``Non-Gaussianities from ekpyrotic collapse with multiple fields,''
  JCAP {\bf 0711}, 024 (2007)
  [arXiv:0708.4321 [hep-th]].

\end{thebibliography}
\end{document}